\documentclass[12pt]{article}
\usepackage{axodraw,bbold}

\catcode`@=12
\begin{document}
\thispagestyle{empty}
\begin{flushright} 
UCRHEP-T425\\ 
December 2006\
\end{flushright}
\vspace{0.5in}
\begin{center}
{\LARGE \bf Application of Finite Groups to Neutrino\\ Mass Matrices\\}
\vspace{1.5in}
{\bf Ernest Ma\\}
\vspace{0.2in}
{\sl Physics and Astronomy Department\\ University of California, Riverside\\ 
Riverside, California 92521, USA \\}
\vspace{1.5in}
\end{center}

\begin{abstract}\
Recent progress in the application of finite groups to neutrino mass 
matrices is reviewed, with special emphasis on the tetrahedral symmetry 
$A_4$.
\end{abstract}

\vspace{0.5in}

\centerline{Talk at VI-Silafae, Puerto Vallarta, November 2006.}

\newpage
\baselineskip 24pt

\section{Introduction}

Using present data from neutrino oscillations, the $3 \times 3$ neutrino
mixing matrix is largely determined, together with the two mass-squared
differences \cite{data}.  In the Standard Model of particle interactions,
there are 3 lepton families.  The charged-lepton mass matrix linking
left-handed $(e, \mu, \tau)$ to their right-handed counterparts is in
general arbitrary, but may always be diagonalized by 2 unitary
transformations:
\begin{equation}
{M}_l = U^l_L \pmatrix{m_e & 0 & 0 \cr 0 & m_\mu & 0 \cr 0 & 0 & m_\tau}
(U^l_R)^\dagger.
\end{equation}
Similarly, the neutrino mass matrix may also be diagonalized by 2 unitary
transformations if it is Dirac:
\begin{equation}
{M}^D_\nu = U^\nu_L \pmatrix{m_1 & 0 & 0 \cr 0 & m_2 & 0 \cr 0 & 0 &
m_3} (U^\nu_R)^\dagger,
\end{equation}
or by just 1 unitary transformation if it is Majorana:
\begin{equation}
{M}^M_\nu = U^\nu_L \pmatrix{m_1 & 0 & 0 \cr 0 & m_2 & 0 \cr 0 & 0 &
m_3} (U^\nu_L)^T.
\end{equation}
Notice that whereas the charged leptons have individual names, the
neutrinos are only labeled as $1,2,3$, waiting to be named.
The observed neutrino mixing matrix is the mismatch between
$U^l_L$ and $U^\nu_L$, i.e.
\begin{eqnarray}
U_{l\nu} = (U^l_L)^\dagger U^\nu_L \simeq \pmatrix{0.83 & 0.56 & <0.2
\cr -0.39 & 0.59 & -0.71 \cr -0.39 & 0.59 & 0.71} \simeq \pmatrix{\sqrt{2/3}
& 1/\sqrt{3} & 0 \cr -1/\sqrt{6} & 1/\sqrt{3} & -1/\sqrt{2} \cr -1/\sqrt{6}
& 1/\sqrt{3} & 1/\sqrt{2}}.
\end{eqnarray}
This approximate pattern has been dubbed tribimaximal by Harrison, Perkins,
and Scott \cite{hps}.  Notice that the 3 vertical columns are evocative
of the mesons $(\eta_8,\eta_1,\pi^0)$ in their $SU(3)$ decompositions.

Historically, once the third lepton $\tau$ was established, it was 
speculated by Cabibbo \cite{c78} and Wolfenstein \cite{w78} that
\begin{equation}
U^{CW}_{l\nu} = {1 \over \sqrt 3} \pmatrix{1 & 1 & 1 \cr 1 & \omega & 
\omega^2 \cr 1 & \omega^2 & \omega},
\end{equation}
where $\omega = \exp(2\pi i/3) = -1/2 + i \sqrt{3}/2$.  Note now
\begin{equation}
U^{HPS}_{l\nu} = (U^{CW}_{l\nu})^\dagger \pmatrix{1 & 0 & 0 \cr 1 & 1/\sqrt 2 
& -1/\sqrt 2 \cr 0 & 1/\sqrt 2 & 1/\sqrt 2} \pmatrix{0 & 1 & 0 \cr 
1 & 0 & 0 \cr 0 & 0 & i}.
\end{equation}
Comparing this to Eq.~(4), it tells us that if $U^l_L$ is in fact 
$U^{CW}_{l\nu}$, then $U^{HPS}_{l\nu}$ can be obtained if maximal 
mixing occurs in the $2-3$ submatrix of $M_\nu$.

How can $U^{HPS}_{l\nu}$ be derived from a symmetry? The
difficulty comes from the fact that any symmetry defined in the basis
$(\nu_e,\nu_\mu,\nu_\tau)$ is automatically applicable to $(e,\mu,\tau)$
in the complete Lagrangian.  To do so, usually one assumes the canonical
seesaw mechanism and studies the Majorana neutrino mass matrix
\begin{equation}
{M}_\nu = -{M}^D_\nu {M}_N^{-1} ({M}^D_\nu)^T
\end{equation}
in the basis where ${M}_l$ is diagonal; but the symmetry apparent
in ${M}_\nu$ (such as $\nu_\mu - \nu_\tau$ interchange) is often 
incompatible with a diagonal ${M}_l$ with 3 very different eigenvalues. 
Obviously a more sophisticated approach is needed.  To obtain 
$U^{HPS}_{l\nu}$, the non-Abelian discrete symmetry $A_4$ turns out to be 
very useful.  In this talk, I will focus mainly on this approach, but 
first I will discuss $S_3$ which is the smallest non-Abelian finite group.  
I will also mention $S_4$ and $\Delta(27)$ at the end.

\section{Permutation Symmetry $S_3$}

$S_3$ is the permutation group of 3 objects, which is also the symmetry 
group of the equilateral triangle.  It has 6 elements divided into 3 
equivalence classes, with the irreducible representations \underline{1}, 
$\underline{1}'$, and \underline{2}, and the multiplication rule
$\underline{2} \times \underline{2} + \underline{1} + \underline{1}' + 
\underline{2}$. Its character table is given below.

\begin{table}[htb]
\centering
\caption{Character table of $S_3$.}
\begin{tabular}{cccccc}
\hline
class&$n$&$h$&$\chi_1$&$\chi_{1'}$&$\chi_2$\\
\hline
$C_1$&1&1&1&1&2\\
$C_2$&2&3&1&1&$-1$\\
$C_3$&3&2&1&$-1$&0\\
\hline
\end{tabular}
\end{table}

\noindent Let me discuss briefly 4 recent $S_3$ models.
\begin{itemize}

\item{Kubo, Mondragon, Mondragon, and Rodriguez-Jauregui, 
\cite{kmmr03} (recently updated by Felix, Mondragon, Mondragon, and 
Peinado \cite{fmmp06}): The symmetry used is actually $S_3 \times Z_2$, 
with the assignments
\begin{equation}
(\nu,l), ~ l^c, ~ N, ~ (\phi^+,\phi^0) \sim \underline{1} + \underline{2},
\end{equation}
and $v_1=v_2$.  The $Z_2$ symmetry serves to eliminate 4 Yukawa couplings 
otherwise allowed by $S_3$, resulting in an inverted ordering of neutrino 
masses with
\begin{equation}
\theta_{23} \simeq \pi/4, ~~~ \theta_{13} \simeq 0.0034, ~~~ m_{ee} \simeq 
0.05~{\rm eV},
\end{equation}
where $m_{ee}$ is the effective Majorana neutrino mass measured in 
neutrinoless double beta decay. This model relates $\theta_{13}$ to the 
ratio $m_e/m_\mu$.}

\item{Chen, Frigerio, and Ma \cite{cfm04}: The symmetry here is $S_3$ 
only, with the assignments
\begin{equation}
(\nu,l) \sim \underline{1} + \underline{2}, ~~~ l^c \sim \underline{1} + 
\underline{1} + \underline{1}', ~~~ (\phi^+,\phi^0) \sim \underline{1} + 
\underline{2}, ~~~ (\xi^{++},\xi^+,\xi^0) \sim \underline{2},
\end{equation}
and $v_1=v_2$ but $u_1 \neq u_2$.  This results in a normal ordering of 
neutrino masses with
\begin{equation}
\theta_{23} \simeq \pi/4, ~~~ 0.008 < \theta_{13} < 0.032, ~~~ m_{ee} < 0.01 
~{\rm eV}.
\end{equation}
This model relates $\theta_{13}$ to $\theta_{12}$ and the ratio 
$\Delta m^2_{sol}/\Delta m^2_{atm}$.}

\item{Grimus and Lavoura \cite{gl05}: The symmetry is $S_3 \times Z_2$, with 
the assignments
\begin{eqnarray}
&& (\nu,l) \sim (\underline{1},+) + (\underline{2},+), ~~~ l^c \sim 
(\underline{1},-) + (\underline{2},+), ~~~ N \sim (\underline{1},-) + 
(\underline{2},-), \nonumber \\ && (\phi^+,\phi^0) \sim (\underline{1},-) + 
(\underline{1},+) + (\underline{1}',+), ~~~ (\chi,\chi^*) \sim 
(\underline{2},+),
\end{eqnarray}
and $\langle \chi \rangle^3 =$ real, resulting in a diagonal $M_l$ and a 
$\mu-\tau$ symmetric $M_\nu$, i.e. $\theta_{23}=\pi/4$ and $\theta_{13}=0$,
whereas $m_{ee}$ is not predicted.}

\item{Mohapatra, Nasri, and Yu \cite{mny06}: The symmetry $S_3$ is extended 
to include 3 $Z_3$ transformations which do not commute with $S_3$, so it is 
not really $S_3$.  For $M_\nu$, the assignments are
\begin{equation}
(\nu,l) \sim \underline{1} + \underline{2}, ~~~ N \sim \underline{1}' + 
\underline{2}, ~~~ (\phi^+,\phi^0) \sim \underline{1}, ~~ (\xi^{++},\xi^+,
\xi^)) \sim \underline{1},
\end{equation}
but the extended $S_3$ is grossly broken by $M_N$ in a very special way, 
resulting then in the tribimaximal form of $M_\nu$.  This is not what I 
would consider a {\it bona fide} derivation of $U^{HPS}_{l\nu}$.}

\end{itemize}

\section{Tetrahedral Symmetry $A_4$}

For 3 families, we should look for a group with a \underline{3}
representation, the simplest of which is $A_4$, the group of the even
permutation of 4 objects, which is also the symmetry group of the
tetrahedron.  The tetrahedron is one of five perfect geometric solids known 
to the ancient Greeks.  In order to match them to the 4 elements (fire, air,
earth, and water) already known, Plato invented a fifth (quintessence)
as that which pervades the cosmos and presumably holds it together.
\begin{table}[htb]
\centering
\caption{Perfect geometric solids in 3 dimensions.}
\begin{tabular}{ccccc}
\hline
solid&faces&vertices&Plato&group\\
\hline
tetrahedron&4&4&fire&$A_4$\\
octahedron&8&6&air&$S_4$\\
cube&6&8&earth&$S_4$\\
icosahedron&20&12&water&$A_5$\\
dodecahedron&12&20&quintessence&$A_5$\\
\hline
\end{tabular}
\end{table}
In terms of symmetry, since a cube (hexahedron) may be embedded inside 
an octahedron and vice versa, the two must have the same group structure 
and are thus dual to each other.  The same holds for the icosahedron and 
dodecahedron.  The tetrahedron is self-dual.
For amusement, compare this first theory of everything to 
today's contender, i.e. string theory.  (A) There are 5 consistent string 
theories in 10 dimensions. (B) Type I is dual to Heterotic $SO(32)$, Type IIA
is dual to Heterotic $E_8 \times E_8$, and Type IIB is self-dual.

$A_4$ has 12 elements divided into 4 equivalence classes, with the 
irreducible representations \underline{1}, $\underline{1}'$, 
$\underline{1}''$, and \underline{3}, and the fundamental multiplication 
rule
\begin{eqnarray}
\underline{3} \times \underline{3} &=& \underline{1}(11+22+33) +
\underline{1}'(11+\omega^222+\omega33) + \underline{1}''
(11+\omega22+\omega^233) \nonumber \\ &+& \underline{3}(23,31,12) +
\underline{3}(32,13,21).
\end{eqnarray}
Its character table is given below, 
\begin{table}[htb]
\centering
\caption{Character table of $A_4$.}
\begin{tabular}{ccccccc}
\hline
class&$n$&$h$&$\chi_1$&$\chi_{1'}$&$\chi_{1''}$&$\chi_3$\\
\hline
$C_1$&1&1&1&1&1&3\\
$C_2$&4&3&1&$\omega$&$\omega^2$&0\\
$C_3$&4&3&1&$\omega^2$&$\omega$&0\\
$C_4$&3&2&1&1&1&--1\\
\hline
\end{tabular}
\end{table}
where $\omega = \exp (2\pi i/ 3) = -1/2 + i\sqrt 3/2$ is exactly what we 
saw before in Eq.~(5). Note that $\underline{3} \times \underline{3} 
\times \underline{3} =
\underline{1}$ is possible in $A_4$, i.e. $a_1 b_2 c_3 +$ permutations,
and $\underline{2} \times \underline{2} \times \underline{2} = \underline{1}$
is possible in $S_3$, i.e. $a_1 b_1 c_1 + a_2 b_2 c_2$.

Other useful sets of finite groups are subgroups of $SU(3)$.  The series 
$\Delta(3n^2)$ has $\Delta(3) \equiv Z_3$, $\Delta(12) \equiv A_4$, 
$\Delta(27)$, etc.  The series $\Delta(3n^2-3)$ has $\Delta(9) \equiv 
Z_3 \times Z_3$, $\Delta(24) \equiv S_4$, etc.
\begin{table}[htb]
\centering
\caption{Representations of $SU(3)$ and its subgroups.}
\begin{tabular}{cccc}
\hline
$SU(3)$&$A_4$&$S_4$&$\Delta(27)$\\
\hline
1&1&1&$1_1$\\
3&3&$3'$&3\\
$\bar{3}$&3&$3'$&$\bar{3}$\\
6&$1+1'+1''+3$&1+2+3&$\bar{3}+\bar{3}$\\
8&$1'+1''+3+3$&$2+3+3'$&$\sum_{i=2,9} 1_i$\\
10&$1+3+3+3$&$1'+3'+3'+3'$&$1_1 + \sum_{i=1,9} 1_i$\\
\hline
\end{tabular}
\end{table}

Using $A_4$, there are two ways to obtain $U^{CW}_{l\nu}$ as the unitary 
matrix which diagonalizes $M_l$: (I) the original proposal of Ma and 
Rajasekaran \cite{mr01} and (II) the recent one by Ma \cite{m06-1}.

\noindent {\bf (I)} Let $(\nu_i,l_i) \sim \underline{3}$, $l^c_i \sim 
\underline{1}, \underline{1}', \underline{1}''$, then with 
$(\phi_i^0,\phi_i^-) \sim \underline{3}$,
\begin{eqnarray}
{M}_l &=& \pmatrix{h_1v_1 & h_2v_1 & h_3v_1 \cr h_1 v_2 & h_2 v_2 \omega
& h_3 v_2 \omega^2 \cr h_1 v_3 & h_2 v_3 \omega^2 & h_3 v_3 \omega} 
\nonumber \\ &=&
{1 \over \sqrt 3} \pmatrix{1 & 1 & 1 \cr 1 & \omega & \omega^2
\cr 1 & \omega^2 & \omega} \pmatrix{h_1 & 0 & 0 \cr 0 & h_2 & 0 \cr 0 & 0
& h_3} \sqrt{3} v,
\end{eqnarray}
for $v_1=v_2=v_3=v$.\\

\noindent {\bf (II)} Let $(\nu_o,l_i) \sim \underline{3}$, $l^c_i \sim 
\underline{3}$, then with $(\phi_i^0,\phi_i^-) \sim \underline{1}, 
\underline{3}$,
\begin{eqnarray}
M_l &=& \pmatrix{h_0v_0 & h_1v_3 & h_2v_2 \cr h_2 v_3 & h_0 v_0 
& h_1 v_1 \cr h_1 v_2 & h_2 v_1 & h_0 v_0} 
\nonumber \\ &=&
{1 \over \sqrt 3} \pmatrix{1 & 1 & 1 \cr 1 & \omega & \omega^2
\cr 1 & \omega^2 & \omega} \pmatrix{m_e & 0 & 0 \cr 0 & m_\mu & 0 \cr 0 & 0
& m_\tau} {1 \over \sqrt 3} \pmatrix{1 & 1 & 1 \cr 1 & \omega^2 & \omega
\cr 1 & \omega & \omega^2},
\end{eqnarray}
for $v_1=v_2=v_3$ with $m_e = h_0 v_0 + (h_1+h_2) v$, $m_\mu = h_0 v_0 + 
(h_1 \omega + h_2 \omega^2) v$, and $m_\tau = h_0 v_0 + (h_1 \omega^2 + 
h_2 \omega) v$.\\

\noindent In either case, $U^{CW}_{l\nu}$ has been derived.  Each allows 
arbitrary values of the charged-lepton masses, and yet retains a symmetry 
for us to consider $M_\nu$.  Let
\begin{equation}
{M}_\nu = \pmatrix{a+b+c & f & e \cr f & a + b\omega + c\omega^2 & d \cr
e & d & a+b\omega^2+c\omega}
\end{equation}
be the Majorana neutrino mass matrix in question.  Under $A_4$, 
$a$ comes from $\underline{1}$, $b$ from $\underline{1}'$, 
$c$ from $\underline{1}''$, and $(d,e,f)$ from $\underline{3}$. 
Since there are 6 free parameters, this is the most general symmetric 
mass matrix.  To proceed further, these 6 parameters must be 
restricted.

\section{Selected $A_4$ Models}

Using {\bf (I)}, the first two proposed $A_4$ models start with only 
$a \neq 0$, yielding thus 3 degenerate neutrino masses.  In Ma and 
Rajasekaran \cite{mr01}, the degeneracy is broken sofly by $N_iN_j$ terms, 
allowing $b,c,d,e,f$ to be nonzero.  In Babu, Ma, and Valle \cite{bmv03}, 
the degeneracy is broken radiatively through flavor-changing supersymmetric 
scalar lepton mass terms.  In both cases, $\theta_{23} \simeq \pi/4$ is 
predicted.  In the latter, maximal CP violation in $U_{l\nu}$ is also 
predicted. Consider the case $b=c$ and $e=f=0$ \cite{m04}, then
\begin{equation}
{M}_\nu = \pmatrix{a+2b & 0 & 0 \cr 0 &
a-b & d \cr 0 & d & a-b},
\end{equation}
which is diagonalized by
\begin{equation}
\pmatrix{1 & 0 & 0 \cr 1 & 1/\sqrt 2 
& -1/\sqrt 2 \cr 0 & 1/\sqrt 2 & 1/\sqrt 2} \pmatrix{0 & 1 & 0 \cr 
1 & 0 & 0 \cr 0 & 0 & i},
\end{equation}
with eigenvalues $a-b+d$, $a+2b$, and $-a+b+d$.  Comparing this with 
Eq.~(6), we see that tribimaximal mixing has been achieved.  However, 
since $\underline{1}'$ and $\underline{1}''$ are unrelated in $A_4$, 
$b=c$ is rather {\it ad hoc}.  A very clever solution by Altarelli and 
Feruglio \cite{af05-1,af05-2} is to eliminate both, then $b=c=0$ naturally.  
This results in a normal ordering of neutrino masses with the prediction 
\cite{m05-1}
\begin{equation}
|m_{\nu_e}|^2 \simeq |m_{ee}|^2 + \Delta m^2_{atm}/9.
\end{equation}
A closely related model by Babu and He \cite{bh05} has $e=f=0$, $b=c$, and 
$d^2=3b(b-a)$.  Here both normal and inverted ordering of neutrino masses 
are allowed.  The technical challenge in this common approach is to break 
$A_4$ spontaneously along two incompatible directions: (1,1,1) and (1,0,0).  
One recent proposal \cite{m06-2} is to add $Z_3$ in a supersymmetric model, 
with singlets carrying the $A_4$ symmetry at a high scale, and require the 
breaking of $A_4$ without breaking the supersymmetry.

As for possible deviations from tribimaximal mixing, although $b \neq c$ 
would allow $U_{e3}$ to be different from zero, the assumption $e=f=0$ 
means that $\nu_2 = (\nu_e + \nu_\mu + \nu_\tau)/\sqrt 3$ remains an 
eigenstate.  The experimental bound $|U_{e3}| < 0.16$ then implies \cite{m04} 
$0.5 < \tan^2 \theta_{12} < 0.52$, whereas experimentally, $\tan^2 \theta_{12} 
= 0.45 \pm 0.05$.\\

\noindent {\bf (III)} A third $A_4$ scenario \cite{hmvv05} is to have 
$(\nu_i,l_i) \sim \underline{3}$, $l^c_i \sim \underline{3}$, 
but with $(\phi^0_i,\phi^-_i) \sim \underline{1}, \underline{1}', 
\underline{1}''$. The charged-lepton mass matrix is now diagonal 
and ${M}_\nu^{(e,\mu,\tau)}= {M}_\nu$ already.  Using again Eq.~(17) but 
with $d=e=f$,
\begin{equation}
{M}_\nu = \pmatrix{a+b+c & d & d \cr d & a+b\omega+c\omega^2 & d \cr
d & d & a+b\omega^2+c\omega}.
\end{equation}
Assume $b=c$ and rotate to the basis $[\nu_e,(\nu_\mu+\nu_\tau)/\sqrt 2,
(-\nu_\mu+\nu_\tau)/\sqrt 2]$, then
\begin{equation}
{M}_\nu = \pmatrix{a+2b & \sqrt 2 d & 0 \cr \sqrt 2 d & a-b+d & 0 \cr
0 & 0 & a-b-d},
\end{equation}
i.e. maximal $\nu_\mu - \nu_\tau$ mixing and $U_{e3}=0$.  The solar mixing
angle is now given by $\tan 2 \theta_{12} = 2\sqrt 2 d/(d-3b)$.  For
$b << d$, $\tan 2 \theta_{12} \to 2\sqrt 2$, i.e. $\tan^2 \theta_{12} \to
1/2$, but $\Delta m^2_{sol} << \Delta m^2_{atm}$ implies $2a+b+d \to 0$, so
that $\Delta m^2_{atm} \to 6bd \to 0$ as well.  Therefore, $b \neq 0$ is
required, and $\tan^2 \theta_{12} \neq 1/2$, but should be close to it,
because $b=0$ enhances the symmetry of ${M}_\nu$ from $Z_2$ to $S_3$.
Here $\tan^2 \theta_{12} < 1/2$ implies inverted ordering and
$\tan^2 \theta_{12} > 1/2$ implies normal ordering.

\section{$S_4$ and $\Delta(27)$}

In the above (III) application of $A_4$, approximate tribimaximal mixing 
involves the {\it ad hoc} assumption $b=c$.  This problem is overcome by 
using $S_4$ in a supersymmetric seesaw model \cite{m05-2}, yielding the
result
\begin{equation}
{M}_\nu(S_4) = \pmatrix{a+2b & e & e \cr e & a-b & d \cr
e & d & a-b}.
\end{equation}
Here $b=0$ and $d=e$ are related limits.  A more recent proposal \cite{m06-3} 
uses $\Delta(27)$, resulting in
\begin{equation}
{M}_\nu(\Delta(27)) = \pmatrix{fa & c & b \cr c & fb & a \cr
b & a & fc}.
\end{equation}

\noindent The permutation group of 4
objects is $S_4$.  It contains both $S_3$ and $A_4$.  It is also the
symmetry group of the hexahedron (cube) and the octahedron.  It has 24 
elements divided into 5 equivalence classes, with 5 irreducible 
representations $\underline{1},\underline{1}',\underline{2},\underline{3},
\underline{3}'$. The fundamental multiplication rules are
\begin{eqnarray}
\underline{3} \times \underline{3} &=& \underline{1}(11+22+33) +
\underline{2}(11+\omega^222+\omega33,11+\omega22+\omega^233) \nonumber \\
&+& \underline{3}(23+32,31+13,12+21) + \underline{3}'(23-32,31-13,12-21),\\
\underline{3}' \times \underline{3}' &=& \underline{1} +
\underline{2} + \underline{3}_S + \underline{3}'_A, ~~~~~~ 
\underline{3} \times \underline{3}' ~=~ \underline{1}' +
\underline{2} + \underline{3}'_S + \underline{3}_A.
\end{eqnarray}
Note that both $\underline{3} \times \underline{3} \times \underline{3} =
\underline{1}$ and $\underline{2} \times \underline{2} \times \underline{2}
= \underline{1}$ are possible in $S_4$.
Let $(\nu_i,l_i),l^c_i,N_i \sim \underline{3}$ under $S_4$.  Assume singlet
superfields $\sigma_{1,2,3} \sim \underline{3}$ and $\zeta_{1,2} \sim
\underline{2}$, then
\begin{equation}
{M}_N = \pmatrix{M_1 & h \langle \sigma_3 \rangle & h \langle \sigma_2 
\rangle \cr h \langle \sigma_3 \rangle & M_2 & h \langle \sigma_1 \rangle \cr 
h \langle \sigma_2 \rangle & h \langle \sigma_1 \rangle & M_3},
\end{equation}
where $M_1 = A+f(\langle \zeta_2 \rangle + \langle \zeta_1 \rangle)$, 
$M_2 = A + f(\langle \zeta_2 \rangle \omega + \langle \zeta_1 \rangle 
\omega^2)$, and $M_3 = A + f(\langle \zeta_2 \rangle \omega^2 + \langle 
\zeta_1 \rangle \omega)$.  The most general $S_4$-invariant superpotential 
of $\sigma$ and $\zeta$ is
given by
\begin{eqnarray}
W &=& M(\sigma_1 \sigma_1 + \sigma_2 \sigma_2 + \sigma_3 \sigma_3) +
\lambda \sigma_1 \sigma_2 \sigma_3 + m \zeta_1 \zeta_2 + \rho(\zeta_1
\zeta_1 \zeta_1 + \zeta_2 \zeta_2 \zeta_2) \nonumber \\
&+& \kappa[(\sigma_1 \sigma_1 + \sigma_2 \sigma_2 \omega + \sigma_3 \sigma_3
\omega^2) \zeta_2 + (\sigma_1 \sigma_1 + \sigma_2 \sigma_2 \omega^2 +
\sigma_3 \sigma_3 \omega) \zeta_1].
\end{eqnarray}
The resulting scalar potential has a minimum at $V=0$ (thus preserving
supersymmetry) only if $\langle \zeta_1 \rangle = \langle \zeta_2 \rangle$
and $\langle \sigma_2 \rangle = \langle \sigma_3 \rangle$, so that
$M_N$ is of the form given by Eq.~(23).  To obtain $M_\nu$ of the same 
form, $M_l$ should be 
diagonal and $M_{\nu N}$ proportional to the identity.  These are both 
possible with $\phi^l_{1,2,3} \sim \underline{1} + \underline{2}$, 
$\phi^N_{1,2,3} \sim \underline{1} + \underline{2}$, but with zero vacuum 
expectation value for $\phi^N_{2,3}$.\\

$\Delta(27)$ has 27 elements divided into 11 equivalence classes.  There are 
9 one-dimensional irreducible representations $\underline{1}_i$ and 2 
three-dimensional ones $\underline{3},\underline{\bar 3}$, with the 
multiplication rules
\begin{equation}
\underline{3} \times \underline{3} = \underline{\bar 3} + \underline{\bar 3} 
+ \underline{\bar 3}, ~~~ \underline{3} \times \underline{\bar 3} = 
\sum_{i=1,9} \underline{1}_i.
\end{equation}
For the product $\underline{3} \times \underline{3} \times \underline{3}$, 
there are 3 invariants: $123+231+312-213-321-132$ which is invariant under 
$SU(3)$, $123+231+312+213+321+132$ which is also invariant under $A_4$, 
and $111+222+333$.  Let $(\nu_i,l_i) \sim \underline{3}$, $l^c_i \sim 
\underline{\bar 3}$, $(\phi^0_i,\phi^-_i) \sim \underline{1}_{1,2,3}$, 
$(\xi^{++}_i,\xi^+_i,\xi^0_i) \sim \underline{3}$, then Eq.~(24) is obtained. 
Again let $b=c$, then two solutions for example are $f=1.1046$ and  
$f=-0.5248$, for both of which $\tan^2 \theta_{12} = 0.45$ and 
$m_{ee} = 0.05$ eV.

\section{Conclusion}

With the application of the non-Abelian discrete symmetry $A_4$,
a plausible theoretical understanding of the tribimaximal form of the 
neutrino mixing matrix has been achieved. Other symmetries such as $S_4$ 
and $\Delta(27)$ are beginning to be studied. They share some of the 
properties of $A_4$ and may help to extend our understanding of possible 
discrete family symmetries, with eventual links to grand unification 
\cite{gut}.

\section*{Acknowledgements} I thank Miguel Perez and the other organizers 
of VI-Silafae for their \underline{great} hospitality and a stimulating 
symposium in Puerto Vallarta.  This work was supported in part by the 
U.~S.~Department of Energy under Grant No.~DE-FG03-94ER40837.


\begin{thebibliography}{99}
\bibitem{data} See for example M. C. Gonzalez-Garcia, these proceedings.
\bibitem{hps} P. F. Harrison, D. H. Perkins, and W. G. Scott, Phys. Lett.
{\bf B530}, 167 (2002).
\bibitem{c78} N. Cabibbo, Phys. Lett. {\bf B72}, 333 (1978).
\bibitem{w78} L. Wolfenstein, Phys. Rev. {\bf D18}, 958 (1978).
\bibitem{kmmr03} J. Kubo, A. Mondragon, M. Mondragon, and 
E. Rodriguez-Jauregui, Prog. Theor. Phys. {\bf 109}, 795 (2003).
\bibitem{fmmp06} O. Felix, A. Mondragon, M. Mondragon, and E. Peinado, 
hep-ph/0610061.
\bibitem{cfm04} S.-L. Chen, M. Frigerio, and E. Ma, Phys. Rev. {\bf D70}, 
073008 (2004).
\bibitem{gl05} W. Grimus and L. Lavoura, JHEP {\bf 0508}, 013 (2005).
\bibitem{mny06} R. N. Mohapatra, S. Nasri, and H.-B. Yu, Phys. Lett. 
{\bf B639}, 318 (2006).
\bibitem{mr01} E. Ma and G. Rajasekaran, Phys. Rev. {\bf D64}, 113012 (2001).
\bibitem{m06-1} E. Ma, hep-ph/0607190.
\bibitem{bmv03} K. S. Babu, E. Ma, and J. W. F. Valle, Phys. Lett.
{\bf B552}, 207 (2003).
\bibitem{m04} E. Ma, Phys. Rev. {\bf D70}, 031901(R) (2004).
\bibitem{af05-1} G. Altarelli and F. Feruglio, Nucl. Phys. {\bf B720}, 64 
(2005).
\bibitem{af05-2} G. Altarelli and F. Feruglio, Nucl. Phys. {\bf B741}, 215 
(2006).
\bibitem{m05-1} E. Ma, Phys. Rev. {\bf D72}, 037301 (2005).
\bibitem{bh05} K. S. Babu and X.-G. He, hep-ph/0507217.
\bibitem{m06-2} E. Ma, hep-ph/0610342.
\bibitem{hmvv05} M. Hirsch, E. Ma, A. Villanova del Moral, and J. W. F.
Valle, Phys. Rev. {\bf D72}, 091301(R) (2005); Erratum-ibid. {\bf D72},
119904 (2005).
\bibitem{m05-2} E. Ma, Phys. Lett. {\bf B632}, 352 (2006).
\bibitem{m06-3} E. Ma, Mod. Phys. Lett. {\bf A21}, 1917 (2006).
\bibitem{gut} C. Hagedorn, M. Lindner, and R. N. Mohapatra, JHEP {\bf 0606}, 
042 (2006); Y. Cai and H.-B. Yu, hep-ph/0608022; I. de Medeiros Varzielas, 
S. F. King, and G. G. Ross, hep-ph/0607054.

\end{thebibliography}
\end{document}